# Analytical Physics-Based Modeling of Electron Channel Density in Nanosheet and Nanowire Transistors

G. I. Zebrev, D. S. Malich

*Abstract*— We propose a general physics-based approach for an accurate analytical calculation of the channel charge density in field-effect transistors as functions of the external gate biases. This approach is based on a consistent consideration of basic electrostatic equation as a balance of electric and chemical potentials which allows us to obtain in a unified way the explicit analytic expressions continuously describing the subthreshold and above threshold regions in nanosheet (symmetric and asymmetric) and nanowire FETs. Two conceptually different definitions of phenomenological threshold voltage are consistently introduced and discussed.

## I. Introduction

THE key advantage of the multi-gate transistors, including FinFETs, nanowire, and nanosheet transistors (GAA FETs), is that their superior electrostatics can effectively suppress the short-channel effects [1, 2]. Another distinctive feature of such transistors is their channels are the thin conductive nanolayer (or nanowire) of slightly doped semiconductor which increases mobility and minimizes the random dopant and stochastic trapped charge fluctuations [3]. The electrostatics of solid-state devices determines their functionality and performance. Electrostatics (Poisson's equation) always works in conjunction with the carriers' statistics and the devices' geometric configurations since the carriers' spatial distributions depend in a self-consistent manner on the potential distributions. The local value of electron density is determined by local chemical potential $\zeta$ or, the same at equilibrium, by electric potential $\varphi$. These potentials are typically not directly measurable and can be controlled only indirectly through the voltages at external contacts. Determining the relationship between the internal physical potentials and external electric biases is one of the main challenges of the device modeling. Generally, we need an accurate analytical de-pendence of the channel electron density on the gate voltage $n_S(V_G)$. In particular, the phenomenological expression $n_S(V_G)$ in the threshold-voltage-based models like BSIM4 [4, 5, 6] and more physical surface-potential-based models [7], have been successfully used in circuit design for many years. The former phenomenological approach omits relevant physics since it is not clear in advance how to modify the phenomenological expression to describe new types of devices (for example, the gate-all-around FETs). The PSP models require as a rule a numerical solution or bulky approximate procedures to obtain $n_S(V_G)$. We intend to show in this paper that the dependence of the charge density for different configurations of FETs can be described over the range of many orders of magnitudes within the framework of an exact analytical solution of the basic electrostatic equation.

## II. Conventional bulk MOSFET case

The electrochemical potential (Fermi energy) in the Si substrate of conventional bulk MOSFETs consists of the sum of chemical $\zeta(x)$ and electrostatic $\varphi(x)$ potentials which is x-independent at any gate voltage

$$\mu = q\zeta(x) - q\varphi(x) = -E_G/2 - q\varphi_F \quad (1)$$

where $\varphi_F$ is the bulk Fermi potential (see Fig.1).

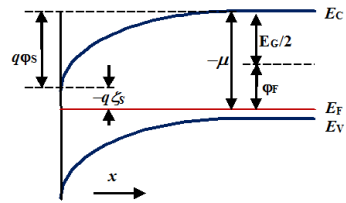

Fig. 1. Energy band diagram of bulk Si-MOSFET.

The basic electrostatic relation can generally be written as a difference between the electrochemical potentials in the gate and Si substrate $E_F(Si) - E_F(gate) = qV_G$ that may be represented in a two-fold way

$$V_G = \varphi_{MS} + \varphi_S + E_{ox}t_{ox} = \varphi_{gate} - \chi_{Si} + \zeta_S + E_{ox}t_{ox}. \quad (2)$$

G. I. Zebrev and D. Malich are with the Department of Micro- and Nanoelectronics of National Research Nuclear University MEPHI, Moscow, Russia, e-mail: gizebrev@mephi.ru.

where $t_{ox}$ is the physical thickness of the gate oxide, $\varphi_{gate}$ and $\chi_{Si}$ are the gate work function and the silicon electron affinity, $\varphi_{MS}$ is the work function difference [8]

$$\varphi_{MS} = \varphi_{gate} - \chi_{Si} - E_g/2q - \varphi_F . \tag{3}$$

where $E_G$ is the Si bandgap, and the electron chemical potential at the interface is defined as $\zeta_S = \varphi_S - E_g/2q - \varphi_F$. In fact, the equilibrium chemical and electrostatic potentials are up to an additive term the same thing in the bulk MOSFETs. A strong electrostatic screening in typically metallic gates makes the potential drop in them negligible. Note that a description with an electrostatic and chemical potential is completely equivalent in equilibrium. When the substrate potential of the bulk MOSFETs is fixed, the surface potential $\varphi_S$ has the meaning of an absolute value that directly determines the concentration of electrons in the channel. Using the electric neutrality condition, one can get

$$\begin{aligned} V_G &= \varphi_{MS} + \varphi_S + E_{ox} t_{ox} = \\ &= \varphi_{MS} + \varphi_S + \frac{q}{C_{ox}}\left[n_S + N_A x_D(\varphi_S) - N_t(\varphi_S)\right], \end{aligned} \tag{4}$$

where, $\varphi_{MS}$ is the work function difference, $C_{ox} = \varepsilon_{ox}/t_{ox}$ is the oxide capacitance, $\varepsilon_{ox}$ is the oxide permittivity, $qN_A x_D(\varphi_S)$ is the depletion layer charge and $N_t(\varphi_S)$ is a sum of the positively and negatively charged interfacial defect densities which is assumed to be positive. Such expressions are ordinarily adapted to practice through the concept of phenomenological threshold voltage, often defined formally as $V_T^* \equiv V_G(\varphi_S = 2\varphi_F)$ [8]. Using an approximation of the uniform depletion layer $C_D$ and the interface trap $C_{it}$ capacitances

$$\begin{aligned} \left[Q_D(\varphi_S) - Q_D(2\varphi_F)\right] + q\left[N_t(2\varphi_F) - N_t(\varphi_S)\right] &\cong \\ &\cong (C_D + C_{it})(\varphi_S - 2\varphi_F), \end{aligned} \tag{5}$$

one can get

$$V_G - V_T^* \cong m(\varphi_S - 2\varphi_F) + \frac{qn_S(\varphi_S)}{C_{ox}}, \tag{6}$$

where $m = 1 + (C_{it} + C_D)/C_{ox}$ is the ideality factor. For the Boltzmann statistics we have

$$n_S \cong N_C t_{inv} \exp(\zeta_S/\varphi_T) = N_A t_{inv} \exp\left(\frac{\varphi_S - 2\varphi_F}{\varphi_T}\right), \tag{7}$$

where $\varphi_T = kT/q$ is the thermal potential, and the effective inversion layer thickness $t_{inv} = \varphi_T/E_{eff}$ depends on the effective electric field $E_{eff} = q(0.5 n_S + N_A x_D)/\varepsilon_S$. Then we have the basic electrostatic equation for the bulk MOSFET in a such form

$$V_G = V_T^* + m\varphi_T \ln\left(\frac{n_S}{N_A t_{inv}}\right) + \frac{qn_S}{C_{ox}}. \tag{8}$$

The electron density $n_S$ varies by many orders of magnitude while the channel thickness $t_{inv}$ alters in a much narrower range, which makes it possible to consider $t_{inv}$ in (8) as approximately constant. Then, the relation (8) can be considered as a transcendental equation for electron density $n_S$. Similar logarithmic terms arose naturally in many papers (e.g., [9, 10, 11], 12, 13).

Despite the specifics of the bulk Si-MOSFET considered, the equation (8) has a rather general character. To show this, consider the transistor swing defined as $S = dV_D/d\ln I_D$ which can be generally represented as

$$S = m\varphi_T + qn_S/C_{ox} . \tag{9}$$

Then, integrating $\int dV_G = \int (S/n_S) dn_S$, we get a general form of the main electrostatic equation for field-effect devices as a sum of chemical and electrical parts

$$V_G - V_{G0} = m\varphi_T \ln\left(\frac{n_S}{n_{S0}}\right) + \frac{q(n_S - n_{S0})}{C_{ox}}, \tag{10}$$

where $V_{G0}$ and $n_{S0} \equiv n_S(V_{G0})$ are arbitrary (not necessarily threshold) reference values. Equation (10) has an accurate analytical solution for electron density

$$qn_S = mC_{ox}\varphi_T W\left[\frac{qn_{S0}}{mC_{ox}\varphi_T} \exp\left(\frac{V_G - V_{G0} + V_{off}}{m\varphi_T}\right)\right], \tag{11}$$

where $W(x)$ is the Lambert function, defined as a solution of the equation $W(xe^x) = x$ [14]. This solution automatically obeys the condition $n_{S0} \equiv n_{S0}(V_{G0})$ for an arbitrary choice of $V_{G0}$. The reference (offset) term $V_{off} = qn_{S0}/C_{ox}$ in (11) is in a sense similar to an empirical fitting parameter $V_{off}$ in BSIM [4]. Optionally, depending on a context of the task, one can choose the midgap voltage $V_{MG}$, the threshold voltage $V_T$, or even the maximum value of the current at the I-V curve as a reference voltage. Equation (11) allows constructing a "minimal" I-V model for a linear operation mode ($V_D \leq \varphi_T$) when $I_D = (W/L)\mu_0 qn_S(V_G)V_D$ ($\mu_0$ is mobility, $W/L$ is the aspect ratio)

$$V_G - V_{G0} = m\varphi_T \ln\left(\frac{I_D}{I_{D0}}\right) + \frac{I_D - I_{D0}}{g_{m0}}, \tag{12}$$

where $g_{m0} \equiv (W/L)\mu_0 C_{ox} V_D$ and $I_{D0} \equiv I_D(V_{G0})$.

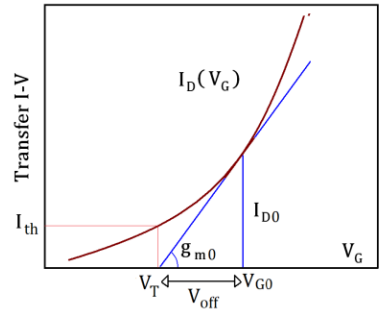

Fig. 2. FET's threshold voltage and current, transconductance and offset voltage V$_{off}$ in one sketch.

Choosing the reference voltage as the point of maximum first derivative [15], or any point on the convex downward portion of the I-V curve, the threshold voltage can be determined as the gate-voltage axis intercept of the linear extrapolation $V_T = V_{G0} - I_{D0}/g_{m0} = V_{G0} - V_{off}$ (see Fig. 2). Then the solution of (12) becomes

$$I_D = g_{m0} m \varphi_T W \left[ \frac{I_{D0}}{g_{m0} m \varphi_T} \exp\left(\frac{V_G - V_T}{m \varphi_T}\right) \right]. \quad (13)$$

The Lambert function has two different asymptotic forms: $W(x) \cong x$ at $x < 1$ and $W(x) \cong \ln x - \ln(\ln(x))$ at $x > 1$. Then we have the following asymptotic expressions for drain current at low $V_{DS}$

$$I_D \cong \begin{cases} I_{D0} \exp\left(\dfrac{V_G - V_T}{m \varphi_T}\right), & V_G \ll V_T, \\ g_{m0}\left(V_G - V_T - m\varphi_T \ln\left(1 + \dfrac{g_{m0}(V_G - V_T)}{I_{D0}}\right)\right), & V_G \gg V_T. \end{cases} \quad (14)$$

The logarithmic term in (14) corresponds to a slight increase in transconductance and uncertainty of $V_T$, typically observed in FETs at moderate overdrives $V_G - V_T$. Thus, (10) describes continuously the subthreshold (exponential) and the strong inversion (linear) regions of the observed channel charge density $n_s$.

### III. SYMMETRIC NANOSHEET FETs

The nanosheet GAA FETs as well as the FinFETs have a flat geometry typically with a thin slightly doped ("intrinsic") silicon body (see Fig. 3a). To obtain the basic electrostatic equation for these structures, it is necessary to solve the 1D Poisson equation for electrostatic potential $\varphi$, which has the following form in a non-degenerate case, which is valid for room temperatures

$$\frac{d^2\varphi}{dx^2} = \frac{q n_0}{\varepsilon_S} \exp\left(\frac{\varphi}{\varphi_T}\right), \quad (15)$$

where $\varepsilon_S$ is the Si permittivity, $n_0 = n(0) = N_C e^{\frac{\zeta_0}{\varphi_T}}$ is the bulk electron concentration in the middle of the silicon nanosheet at $x = 0$ where the chemical potential $\zeta_0 = \zeta(0)$.

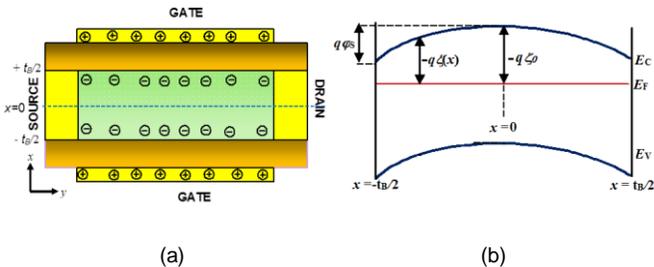

(a)  (b)

Fig. 3. Schematic view (a) and energy band diagram (b) of symmetric nanosheet FET.

Here we consider the case when the Fermi level is in the upper half of the Si bandgap so that the dopant and hole concentrations can be neglected. More specifically, this can be done when the inequality $\zeta_0 > \varphi_T \ln N_A / n_i - E_G / 2q$ is obeyed. Taur et al. [16] and many others used the boundary condi-tions in a form of fixed electric field at the Si-SiO$_2$ interfaces. Such boundary conditions fix the surface charge density in the Si nanolayer, but do not explicitly determine the position of the Fermi level in it, which leads to the absence of a closed analytical solution and, as a consequence, to unnecessary technical difficulties. The boundary conditions for a symmetric connection can be written in another form $\varphi(0) = 0, \; d\varphi/dx|_{x=0} = 0$. The former condition corresponds to an explicit fixed value of the electrochemical potential at $x=0$ $\mu_0/q = \zeta_0 = \varphi_T \ln(n_0/N_C)$ which remains x-independent over the entire thickness of the Si body (see Fig. 2b). Then the exact solution of (14) in the symmetric case has a simple form

$$\varphi(x) = 2\varphi_T \ln\left[\sec\left(\frac{x}{L_D}\right)\right], \quad (16)$$

where the Debye length is defined by an electron concentration at the middle of the Si body

$$L_D = \left(\frac{2\varepsilon_S \varphi_T}{q n_0}\right)^{1/2} = \left(\frac{2\varepsilon_S \varphi_T}{q N_C}\right)^{1/2} e^{-\frac{\zeta_0}{2\varphi_T}} \equiv L_D^{\min} e^{-\frac{\zeta_0}{2\varphi_T}}. \quad (17)$$

The minimum Debye length occurs approximately at the degeneration onset $n_0 \cong N_C = 3.3 \times 10^{19}$ cm$^{-3}$ (Si) and it is equal $L_D^{\min} \cong 1$ nm. The maximum Debye length is limited on top by unintentional doping of the silicon layer $L_D^{\max} = (2\varepsilon_S \varphi_T / q N_A)^{1/2}$ and does not affect the results in any way. The electric potential difference between the middle of the Si body and the Si-dielectric interfaces is given by

$$\varphi_S = 2\varphi_T \ln\left[\sec\left(\frac{t_B}{2L_D}\right)\right]. \quad (18)$$

The total surface electron concentration in the silicon sheet is calculated accurately by the integral over the entire thickness of the Si sheet

$$n_S = n_0 \int_{-t_B/2}^{t_B/2} e^{\frac{\varphi(x)}{\varphi_T}} dx = 2 n_0 L_D \tan\left(\frac{t_B}{2L_D}\right) = \\ = n_0 L_D \left(e^{\frac{\varphi_S}{\varphi_T}} - 1\right)^{1/2} = 2 N_C L_D^{\min} e^{\frac{\zeta_0}{2\varphi_T}} \left(e^{\frac{\varphi_S}{\varphi_T}} - 1\right)^{1/2}. \quad (19)$$

In contrast to the bulk MOSFET case (2), the chemical $\zeta_0$ and electric $\varphi_S$ potentials are physically different things in GAAFETs, defined therein at different points. The former is determined by the electron density $n_0$ and explicitly controlled by the gate bias. The latter has a geometric origin and can be reduced to negligible values in very thin nanosheets.

Fig. 4 shows the profiles of bulk concentrations calculated with an accurate relation (16)

$$n(x) = n_0 e^{\frac{\varphi(x)}{\varphi_T}} = N_C e^{\frac{\zeta_0 + \varphi(x)}{\varphi_T}} \quad (20)$$

For weak inversion mode $\zeta_0 < \varphi_T \ln(2L_D^{\min}/t_B)$, or, equivalently, $t_B/2L_D \ll 1$, we have from (21) an approximately uniform distribution of bulk electron concentration over the nanosheet thickness

$$n_S \cong n_0 t_B = N_C t_B e^{\zeta_0/\varphi_T}. \quad (21)$$

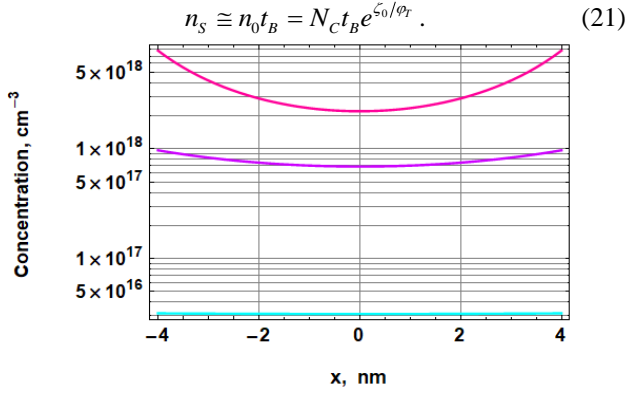

Fig. 4. Electron concentration distributions over a nanolayer ($t_B$ = 8 nm) calculated with (20) for chemical potentials $\zeta_0$ = - 7, -10, and – 18 mV.

Besides, expanding (19) in terms of small parameter $t_B/2L_D$, we find the potential drop over the half thickness of the Si body is much less than a thermal potential

$$\varphi_S \cong \varphi_T \frac{t_B^2}{4L_D^2} = qn_S \frac{t_B}{8\varepsilon_S} < \varphi_T. \quad (22)$$

For the strong inversion regime and/or for thick silicon layer when $t_B/2L_D > 1$ and $\zeta_0 > \varphi_T \ln(2L_D^{min}/t_B) \sim -2\varphi_T$ we have $\varphi_S > \varphi_T$, and

$$n_S \cong 2N_C L_D^{min} e^{\frac{\zeta_0+\varphi_S}{2\varphi_T}}, \quad (23)$$

where $\zeta_0 + \varphi_S$ is a local value of the chemical potential at the interfaces (see Fig. 3(b)). Equation (23) corresponds to a double inversion layer mode when $dn_S/d\zeta \cong n_S/2\varphi_T$ as is the case in bulk MOSFETs.

Neglecting for brevity the interface traps and omitting temporarily an additive reference bias, the basic electrostatic relation for nanosheet transistors can be written in an accurate form

$$V_G = \zeta_0 + \varphi_S(\zeta_0) + \frac{qn_S(\zeta_0)}{2C_{ox}} \quad (24)$$

where $\varphi_S$ and $n_S$ represent the explicit and exact functions (18) and (19) of chemical potential $\zeta_0$ at the middle of the Si nanosheet. Equation (22) allows us to implicitly construct an exact dependence $n_S(\zeta_0)$ vs $V_G(\zeta_0)$ but does not allow obtaining an explicit analytical dependence $n_S(V_G)$. A reasonable approach for obtaining of analytical relation $n_S(V_G)$ is a use of the "thin sheet" ($t_B/2L_D \ll 1$) approximation (21) and (22). Then, the basic electrostatic relation for nanosheet transistors can be written in a difference form with respect to a reference voltage $V_{G0}$.

$$V_G - V_{G0} = \Delta\zeta_0 + \Delta\varphi_S + \frac{q\Delta n_S}{2C_{ox}} \cong \Delta\zeta_0 + \frac{q\Delta n_S}{C_{tot}}, \quad (25)$$

where $\Delta n_S = n_S(V_G) - n_S(V_{G0})$, $\Delta\zeta_0 = \zeta_0(V_G) - \zeta_0(V_{G0})$, and the effective capacitance is defined from (21) as follows

$$\frac{1}{C_{tot}} = \frac{t_{ox}}{2\varepsilon_{ox}} + \frac{t_B}{8\varepsilon_S}, \quad (26)$$

where the first term corresponds to the oxide voltage drop while the second term describes the voltage drop on the half-width of the Si body (22). Taking into account that $\zeta_0(V_G) \cong \varphi_T \ln(n_S(V_G)/N_C t_B)$ in this approximation, and choosing the threshold voltage $V_{th}$ as a reference point, defined, alternative to $V_T$ in (12), by a given level of a threshold concentration $n_S^{th}$ (~ $10^{11}$ cm$^{-2}$), the basic electrostatic equation can be written in a consistent generic form

$$V_G - V_{th} \cong \varphi_T \ln\left(\frac{n_S}{n_S^{th}}\right) + \frac{q(n_S - n_S^{th})}{C_{tot}}, \quad (27)$$

which has an exact solution

$$qn_S = C_{tot}\varphi_T W\left[\frac{qn_S^{th}}{C_{tot}\varphi_T}\exp\left(\frac{V_G - V_{th}}{\varphi_T} + \frac{qn_S^{th}}{C_{tot}\varphi_T}\right)\right], \quad (28)$$

identically satisfying a condition $n_S(V_{th}) = n_S^{th}$.

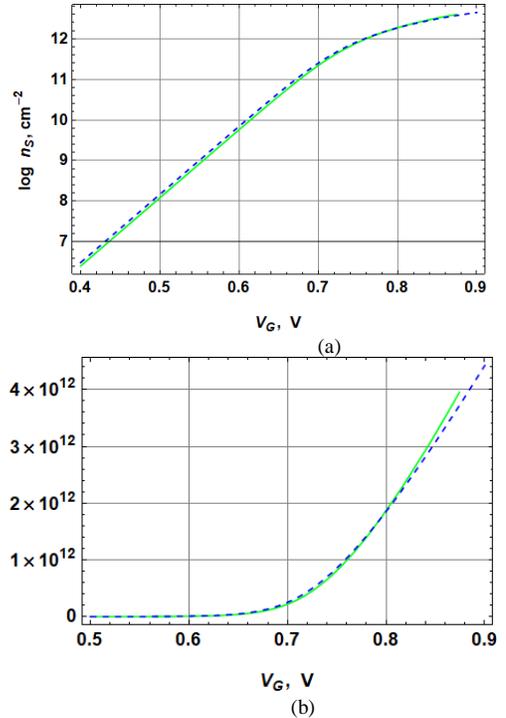

Fig. 5. Comparison of accurate solution (solid lines) and approximate simulation (28) for $n_S(V_G)$ (dashed lines) performed for EOT = 1 nm, $t_B$ = 5 nm in logarithmic (a) and linear (b) scales. The two curves are aligned without distortion by fitting the phenomenological parameter $V_{th}$ and $V_{off}$ in (28).

This approximation is very good in a relevant range of nanosheet thicknesses $t_B < 8$ nm (see Fig. 5). As expected, the approximate and exact expressions agree well in the sub-threshold region or at weak inversion and show a small discrepancy at very strong inversion, where the inhomogeneity of the electron gas distribution becomes noticeable (see Fig. 4). Fig. 6 shows the bulk electron density distribution simulated as a function of gate voltage

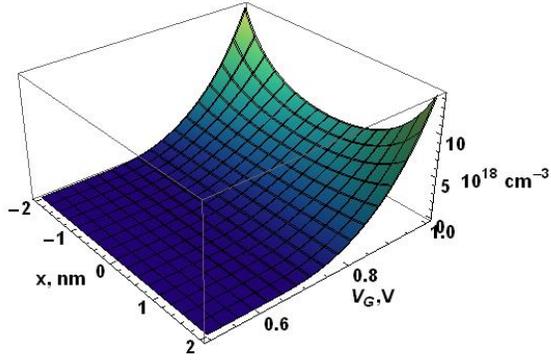

Fig. 6. Bulk electron concentration as a function of coordinate x in nanosheet FET and gate voltage $V_G$, calculated for $t_B$ = 4 nm.

Fig. 7 shows validation of the model using a "minimal" linear I-V model presented in Sec. II.

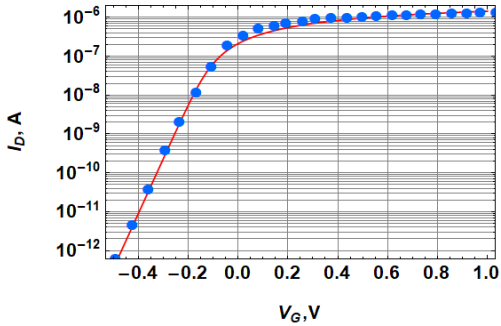

Fig. 7. Transfer I-V curve of the NSFETs [17] ($V_D$ = 50 mV, $t_B$ = 6 nm, $t_{ox}$ (SiO$_2$) = 5 nm, W/L=23nm/22nm) compared to simulation results (a fitting mobility is 350 cm$^2$/V-cm).

It should be emphasized that an approximation $t_B/2L_D \ll 1$ that allows an analytic expression is certainly valid for ultrathin nanosheets and can be used in a modified form even for monolayers with a nonzero band gap.

## IV. ASYMMETRIC DOUBLE-GATE FETs

If different gate-source biases $V_{G1}$ and $V_{G2}$ are applied for an asymmetric double–gate FET, then the position of the minimum potential value (virtual cathode) $x_0$ and the electron concentration $n_0 = n(x_0)$ at this point depend in general on their values. Solving the Poisson's equation (15) in this case has the form

$$\varphi(x) = 2\varphi_T \ln\left[\sec\left(\frac{x - x_0}{L_D}\right)\right], \quad (29)$$

where the Debye length $L_D$ should be determined at the point of the virtual cathode (minimum potential). The total electron concentration per unit area of the silicon body is calculated as in (19)

$$n_S = n_0 L_D \left(\tan\left(\frac{t_B - x_0}{L_D}\right) + \tan\left(\frac{x_0}{L_D}\right)\right) = n_0 L_D \frac{\sin(t_B/L_D)}{\cos[x_0/L_D]\cos[(t_B - x_0)/L_D]}, \quad (30)$$

that coincides with (19) for the symmetric case $x_0 = t_B/2$. The virtual cathode position should generally be determined from the boundary values of the electric field specified by the gate voltages [18]. Here we are interested only in an important special case of a very thin body $t_B \ll L_D$, for that the electron density is practically independent of the virtual cathode position $n_S \cong n_0 t_B$. This is it directly follows from (30). Neglecting for simplicity the voltage drops on the Si sheet we can determine $n_S$ by the basic electrostatic (electric neutrality) equation, which is generalized for asymmetric structure as follows [19]

$$C_1(V_1 - \zeta_0) + C_2(V_2 - \zeta_0) \cong qn_S, \quad (31)$$

where, in the general case, different values of oxide capacitances $C_{1(2)}$ and gate biases $V_{1(2)} = V_{G1(2)} - V_{th}$ are assumed. This equation can be readily reduced to a standard form

$$V_{Geff} = \zeta_0 + \frac{qn_S(\zeta_0)}{C_1 + C_2} = \varphi_T \ln\left(\frac{n_S}{n_S^{th}}\right) + \frac{qn_S}{C_1 + C_2}, \quad (32)$$

where the effective gate voltage is defined as follows

$$V_{Geff} = \frac{C_1 V_1 + C_2 V_2}{C_1 + C_2}. \quad (33)$$

Accordingly, the exact solution of (32) has a form

$$qn_S = (C_1 + C_2)\varphi_T W\left[\frac{V_{off}}{\varphi_T}\exp\left(\frac{V_{Geff} + V_{off}}{\varphi_T}\right)\right], \quad (34)$$

where $V_{off} = qn_S^{th}/(C_1 + C_2)$ and $n_S(V_{Geff} = 0) \equiv n_S^{th}$.

### A. Nanowire FETs

A nanowire FET is another promising type of GAA FETs We assume that the channel (body) of the nanowire transistors has a form of the semiconductor cylinder with a radius $r_B$. Neglecting the quantum confinement effects, the Poisson's equation in cylindrical coordinates is written as

$$\frac{1}{r}\frac{d}{dr}\left(r\frac{d\varphi}{dr}\right) = \frac{qn_0}{\varepsilon_S}\exp\left(\frac{\varphi}{\varphi_T}\right), \quad (35)$$

where $n_0$ is the bulk concentration of electrons on the structure axis. To avoid a cumbersome analysis of the general case, we will use from the very beginning the smallness of the nanowire radius in comparison with the Debye length $r_B < L_D = (2\varphi_T \varepsilon_S/qn_0)^{1/2}$. In this case, the electron distribution in the r. h. s. of (34) can be considered as uniform, and its solution with the zero boundary conditions has a form

$$\varphi(r) \cong \frac{qn_0 r^2}{4\varepsilon_S}. \quad (36)$$

Then the voltage drop across the nanowire radius is written as

$$\varphi_S = \frac{qn_0 r_B^2}{4\varepsilon_S} = \frac{qn_L}{4\pi\varepsilon_S}, \quad (37)$$

where $n_L \cong \pi n_0 r_B^2$ is the linear electron density in a nanowire. The radial distribution of the electric field magnitude in the ring oxide layer $r_B \leq r \leq r_B + t_{ox}$ is calculated from the Gauss law $E_r(r) = qn_L/2\pi\varepsilon_{ox}r$ and potential drop on the oxide is

$$\Delta\varphi_{ox} = \int_{r_B}^{r_B+t_{ox}} E_r(r) dr = \frac{qn_L}{2\pi\varepsilon_{ox}} \ln\left(1 + \frac{t_{ox}}{r_B}\right). \quad (38)$$

Taking into account $n_L \cong \pi r_B^2 N_C \exp(\zeta_0/\varphi_T)$, the basic electrostatic equation for nanowire transistors can be written in the following form

$$V_{GS} = \zeta_0 + \varphi_S + \Delta\varphi_{ox} \cong \varphi_T \ln\left(\frac{n_L}{N_C \pi r_B^2}\right) + \frac{qn_L}{C_{NW}}, \quad (39)$$

where the nanowire specific capacitance per unit length is represented as serial capacitances of body and oxide

$$\frac{1}{C_{NW}} = \frac{1}{4\pi\varepsilon_S} + \frac{\ln(1+t_{ox}/r_B)}{2\pi\varepsilon_{ox}}. \quad (40)$$

Then, the linear electron density is an explicit function of the chemical potential at the center of the nanowire

$$qn_L = C_{NW}\varphi_T W\left[\frac{V_{off}}{\varphi_T}\exp\left(\frac{V_G - V_{G0} + V_{off}}{\varphi_T}\right)\right] \quad (41)$$

where $V_{off} = qn_L(V_{G0})/C_{NW}$.

## V. Limitations and applications

The solution of the main electrostatic equation in the form of the Lambert function is common for non-degenerate systems at finite temperatures where the chemical potential can be represented by a logarithm of carriers' density. The obtained relations $n_S(V_G)$ can be applicable even for monolayers with a nonzero band gap (for example, in molybdenite MoS$_2$ [19]), while they are inapplicable, for example, for graphene. The electrostatic equation for degenerate systems does not necessarily contain a logarithmic term in the actual solution domain, and the Lambert function does not appear in the solution. At the same time, the described methodology for calculating $n_S(V_G)$ is universal for field-effect devices. An important example of degenerate systems is a gapless graphene which allows us to explicitly solve the electrostatic equation in rational functions [20]. We also neglected in this paper some non-trivial quantum effects such as sub-band separation which give rise to kinks and peaks in the capacitance and transconductance [21]. It was shown in [22] that for silicon bodies with thicknesses $t_B > 3$nm the quantum confinement is not significant enough to cause kinks and peaks.